\documentclass[sigconf]{acmart}
\usepackage{url}
\usepackage{csquotes}

\renewenvironment{quote}
  {\list{}{\rightmargin=0.4cm \leftmargin=0.4cm}%
   \item\relax}
  {\endlist}
  
\newcommand{\rr}[1]{\textcolor{black}{#1}}

\newcommand{\datadone}[1]{}

\AtBeginDocument{%
  \providecommand\BibTeX{{%
    \normalfont B\kern-0.5em{\scshape i\kern-0.25em b}\kern-0.8em\TeX}}}

\copyrightyear{2023} 
\acmYear{2023} 
\setcopyright{acmlicensed}\acmConference[ASSETS '23]{The 25th International ACM SIGACCESS Conference on Computers and Accessibility}{October 22--25, 2023}{New York, NY, USA}
\acmBooktitle{The 25th International ACM SIGACCESS Conference on Computers and Accessibility (ASSETS '23), October 22--25, 2023, New York, NY, USA}
\acmPrice{15.00}
\acmDOI{10.1145/3597638.3608385}
\acmISBN{979-8-4007-0220-4/23/10}


\acmConference[ASSETS]{The ACM SIGACCESS Conference on Computers and Accessibility}{Oct 22--25, 2023}{New York, United States}

\acmSubmissionID{2062}
\begin{document}

\title[Day-to-Day Experiences with Conversational Agents after TBI]{Investigating the Day-to-Day Experiences of Users with Traumatic Brain Injury with Conversational Agents}


\author{Yaxin Hu}
\orcid{0000-0003-4462-0140}
\affiliation{%
  \institution{Department of Computer Sciences\\University of Wisconsin--Madison}
  \streetaddress{Department of Computer Sciences, University of Wisconsin--Madison}
  \country{} 
}
\email{yaxin.hu@wisc.edu}

\author{Hajin Lim}
\orcid{0000-0002-4746-2144}
\affiliation{%
  \institution{Department of Communication\\Seoul National University}
  \streetaddress{Department of Communication, Seoul National University, Seoul, South Korea}
  \country{} 
  }
\email{hajin@snu.ac.kr}

\author{Hailey L. Johnson}
\orcid{0000-0003-1310-9948}
\affiliation{
  \institution{Department of Computer Sciences\\University of Wisconsin--Madison}
  \streetaddress{Department of Computer Sciences, University of Wisconsin--Madison}
  \country{} 
}
\email{hljohnson22@wisc.edu}

\author{Josephine M. O'Shaughnessy}
\orcid{0009-0000-3975-8810}
\affiliation{%
  \institution{Department of Communication Sciences and Disorders \\University of Wisconsin--Madison}
  \streetaddress{University of Wisconsin--Madison}
  \country{} 
}
\email{oshaughness2@wisc.edu}

\author{Lisa Kakonge}
\orcid{0000-0003-4164-6716}
\affiliation{%
  \institution{Rehabilitation Science\\McMaster University}
  \streetaddress{Rehabilitation Science, McMaster University}
  \country{} 
}
\email{kakongel@mcmaster.ca}

\author{Lyn S. Turkstra}
\orcid{0000-0002-6948-6921}
\affiliation{%
  \institution{Rehabilitation Science\\McMaster University}
  \streetaddress{Rehabilitation Science, McMaster University}
  \country{} 
}
\email{turkstrl@mcmaster.ca}

\author{Melissa C. Duff}
\orcid{0000-0003-1759-3634}
\affiliation{%
  \institution{Dept. of Hearing \& Speech Sciences\\Vanderbilt University Medical Center}
  \streetaddress{Vanderbilt University Medical Center}
  \country{} 
}
\email{melissa.c.duff@vanderbilt.edu}

\author{Catalina L. Toma}
\orcid{0000-0003-0714-312X}
\affiliation{%
  \institution{Department of Communication Arts\\University of Wisconsin--Madison}
  \streetaddress{Department of Communication Arts, University of Wisconsin-Madison}
  \country{} 
}
\email{ctoma@wisc.edu}

\author{Bilge Mutlu}
\orcid{0000-0002-9456-1495}
\affiliation{%
  \institution{Department of Computer Sciences\\University of Wisconsin--Madison}
  \streetaddress{Department of Computer Sciences, University of Wisconsin--Madison}
  \country{} 
}
\email{bilge@cs.wisc.edu}










\renewcommand{\shortauthors}{Hu et al.}

\begin{abstract}


Traumatic brain injury (TBI) can cause cognitive, communication, and psychological challenges that profoundly limit independence in everyday life. Conversational Agents (CAs) can provide individuals with TBI with cognitive and communication support, although little is known about how they make use of CAs to address injury-related needs. In this study, we gave nine adults with TBI an at-home CA for four weeks to investigate use patterns, challenges, and design requirements, focusing particularly on injury-related use. The findings revealed significant gaps between the current capabilities of CAs and accessibility challenges faced by TBI users. We also identified 14 TBI-related activities that participants engaged in with CAs. We categorized those activities into four groups: mental health, cognitive activities, healthcare and rehabilitation, and routine activities. Design implications focus on accessibility improvements and functional designs of CAs that can better support the day-to-day needs of people with TBI.

\end{abstract}



\maketitle

\begin{figure}[!ht]
    \centering
    \includegraphics[width=\columnwidth]{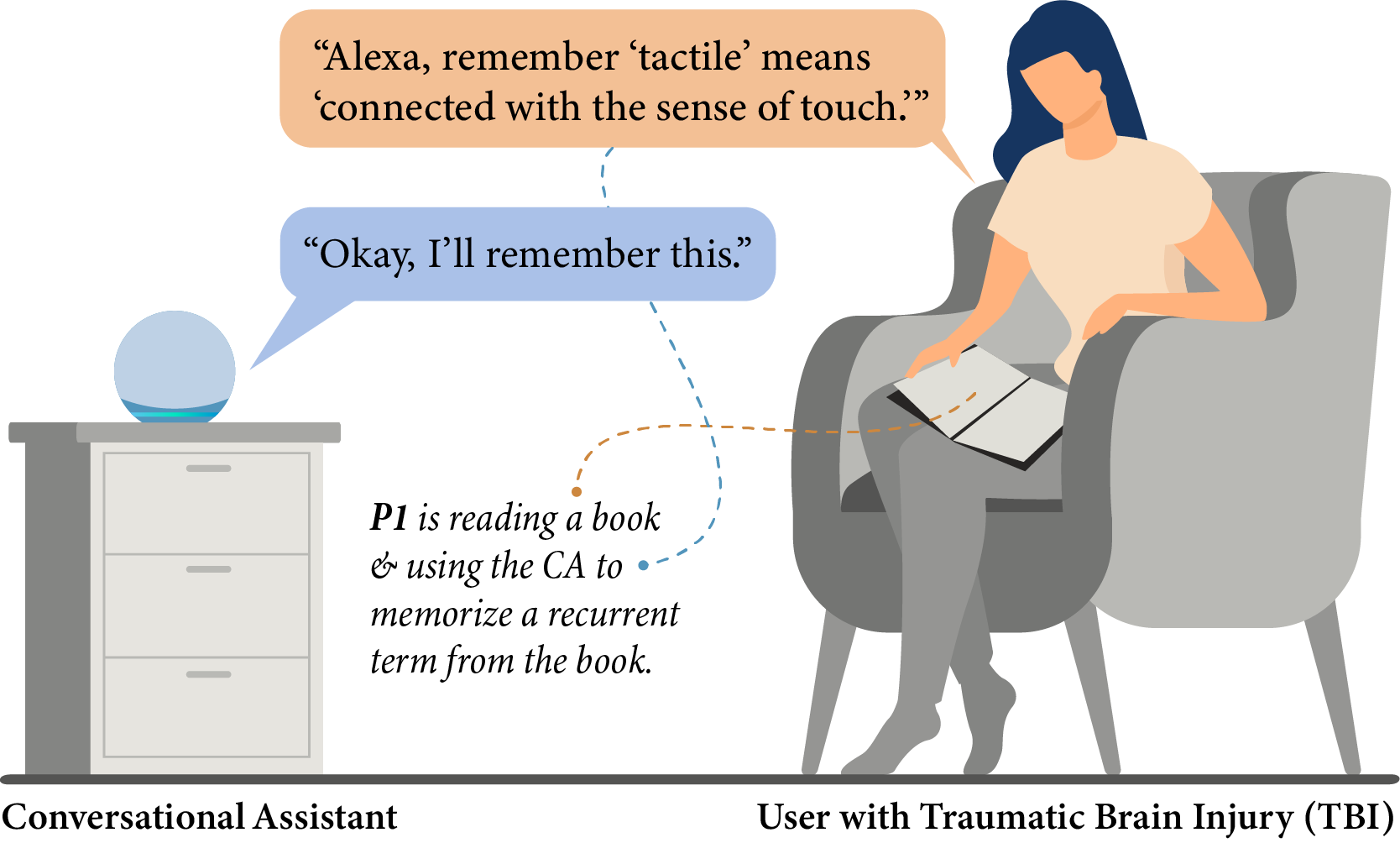}
    \caption{In this paper, we investigate the use of conversational agents (CA) by users with traumatic brain injury (TBI) through a four-week-long in-home study. \rr{Our findings revealed the TBI-related symptoms and challenges this population faces and 14 activities in which participants used CAs to overcome their injury-based challenges. The figure exemplifies one of these activities, \textit{Reading Support}, using data from \textbf{P1}. These activity patterns, along with the gaps between the CA's capabilities and users' injury-based needs, point to design opportunities to support everyday activities of individuals with TBI.}}
    \Description[teaser image]{ Figure 1: The image shows a user with traumatic brain injury interacting with a conversational agent. The activity is annotated as "P1 is reading a book & using the CA to memorize a recurrent term from the book" The dialog between the user and the CA is as follows: The user said:  ``Alexa, remember 'tactile' means 'connected with the sense of touch.''' and the CA responds with ``Okay, I'll remember this.''}
    \label{fig:teaser}
\end{figure}

\clearpage

\section{Introduction}

Conversational Agents (CAs) engage in conversational interactions with people using written or spoken language \cite{allouch2021conversational, car2020conversational}. Advances in natural language processing, speech recognition, and cloud-based services have enabled sophisticated CA systems that can engage in natural conversations with users over an increasingly diverse set of topics and deliver information and services conveniently and effectively. CAs are becoming more prevalent in human environments through integration into devices such as smartphones, cars, speakers, televisions, and wearables \cite{car2020conversational, ahire2022ubiquitous}. Prior work has characterized the general functions of CAs (e.g., music-playing, information retrieval, home integration, and weather information delivery) \cite{bentley2018longterm, Lopatovska2019talkToMeAlexa, ammari2019reallyuse}, how they integrate into people's everyday lives and social interactions \cite{porcheron2018everyday, alex2018inhomeCA, beneteau2020alexafamily}, and the use patterns and challenges associated with interaction with ``conversational'' devices \cite{clark2019goodconversation, grudin2019quest}. Recent advances, such as the emergence of large language models (LLMs), and their integration with CAs \cite{sakirin2023user,jo2023understanding}, point to future CA technologies that offer more sophisticated capabilities and complex interactions.

As an everyday technology, CAs can also serve as \textit{assistive} technologies to help individuals with disabilities overcome barriers to independent living and improve their quality of life \cite{vieira2022disabilities}. Prior research has explored how CAs can support individuals with vision impairments \cite{abdolrahmani2018blindPA, choi2020fast}, older adults \cite{pradhan2020lowtech, kim2021longitudinal}, and users with traumatic brain injury (TBI) \cite{malapaschas2020accessibility}, dementia \cite{moyle2019promise}, and other cognitive impairments \cite{masina2020investigating}. Our work focuses on the use and potential benefits of CA technologies for individuals with TBI as a population with a significant need for social and cognitive support. Although prior work offers a preliminary understanding of the needs and expectations of this population for social and cognitive support and how CA technologies can offer such support, this small body of literature \cite[e.g.,][]{malapaschas2020accessibility,malapaschas2021voice} is based on surveys and interviews based on short-term exposure to the technology. More research is needed to better understand the real-world use of CA technologies by this population, particularly the use patterns that emerge over time, the challenges that users face in using CAs, and the unmet needs of the population by the design of currently available technologies. Such an understanding can inform the design of future CA technologies as well as the development of new therapeutic interventions that integrate these technologies. 

To gain a better understanding of the use patterns, challenges, and unmet needs of individuals with TBI with respect to CAs, we carried out a study with nine adults with moderate-severe TBI in the chronic stage (1 year or more post-injury) who received CAs (Alexa on Amazon Echo Dot) and were asked to use it daily for a period of four weeks in their homes. Participants had reported diverse cognitive, communication, and sensorimotor symptoms and challenges related to TBI as well as changes in their psychological states and ability to manage social situations. Data collection included dialog history from the CA, audio recordings of the context of the user interaction with the CA use captured using a tablet computer, and weekly semi-structured interviews with participants. Our analysis of the data revealed that CAs hold promise to support individuals with TBI across four categories of needs: (1) mental health; (2) cognitive activities; (3) healthcare and rehabilitation; and (4) routine activities. We draw connections between injury-related symptoms and challenges and identify design and accessibility deficiencies as well as opportunities.

The contributions of our work fall under two categories:
\rr{
\begin{itemize}
\item An understanding of what injury-related needs users with TBI sought to address using CAs and how well existing CA capabilities meet these needs based on naturalistic data characterizing how individuals with TBI used CAs;
\item Design recommendations to address gaps between injury-related needs and system capabilities and to meet user expectations from systems that can offer cognitive, emotional, and practical support.
\end{itemize}
}

In the next section (\S\ref{sec:related}), we present related work, focusing on the injury-related symptoms and technology-use challenges of individuals with TBI, the use of CAs as assistive technologies, and field studies of day-to-day CA use. We next present the design of our study (\S\ref{sec:method}) and its findings in two parts: \S\ref{sec:results-part1} reports on injury-related symptoms and accessibility challenges experienced by our participants, and \S\ref{sec:results-part2} presents use patterns emerged during the four-week study period related to these symptoms and challenges. In \S\ref{sec:discussion}, we discuss the implications of both sets of findings on the accessibility design and functional design of CA technologies to support TBI-based needs, and the limitations of our study with pointers to our future work.

\section{Related Work}\label{sec:related}

\rr{Our work draws from literature on the challenges that the adults with TBI face day to day, technology use by this population, and the use, accessibility, and prior studies of CA technologies. We provide brief summaries of relevant research below.}

\subsection{TBI-related Challenges}
TBI is an acquired injury to the brain resulting from an external bodily force, typically from road traffic injuries and falls \cite{iaccarino2018epidemiology}. Effects of TBI are heterogeneous, depending in part on the severity of the initial injury \cite{covington2021heterogeneity}, but for moderate-severe brain injuries, sequelae commonly include impairments in sensorimotor, psychological, and cognitive functions \cite{mazaux1997long}. Cognitive impairments, in particular, can have profound effects on everyday life. Common impairments include challenges in prospective memory; the ability to remember to carry out future tasks \cite{mathias2005prospective}. Working memory challenges also are ubiquitous after TBI \cite{lannoo1998subjective, rochat2013inhibition}, as are impairments in declarative learning and recall \cite{spitz2012association, mazaux1997long, vakil2005effect}. Impairments in executive functions and attention also are common, including challenges with self-control and flexible thinking \cite{mazaux1997long}. Cognitive challenges, in turn, can translate into communication challenges, particularly challenges in using appropriate language in social contexts, and keeping pace with complex conversations \cite{togher2014incog}. All of the above contribute to significant social challenges for adults with moderate-severe TBI \cite{mazaux1997long,doig2001patterns, hoofien2001traumatic,engberg2004psychosocial, wise2010impact,jourdan2016comprehensive}. Social challenges often combine with limitations in family support, reduced friendship networks, and barriers to accessing social activities, leading many people with TBI to experience social isolation and loneliness \cite{levack2010experience}. People with TBI may find cognitive and social support through the use of CAs. However, it is still unknown if CAs have the features needed to provide the necessary support or if CAs are accessible enough for adults with TBI.

\subsection{CAs as Assistive Technologies}

Prior work has shown the benefits of CA usage in a variety of health domains. For individuals with mental health symptoms, regular interaction with CAs have shown to increase meditation frequency and duration \cite{hudlicka2013virtual}, increase motivation to continue behavioral change interventions \cite{lisetti2013can}, and reduce depression symptoms \cite{fitzpatrick2017delivering, pinto2013avatar}. \citet{lucas2017reporting} found that people with combat-related mental health conditions (e.g., post-traumatic stress disorder) were more likely to disclose their symptoms to a CA compared to a health professional or an anonymous health assessment. For people with cognitive impairments, CAs have been shown to have positive usability in people with memory problems \cite{boumans2022voice}, empower both the individual and their care partners \cite{zubatiy2021empowering}, and potentially serve as a monitor of cognitive impairments through entertainment \cite{de2022automatic}. \citet{rampioni2021embodied} reviewed the literature on CAs' usability for people with dementia and found that individuals were naturally engaged and attentive when using CAs. For people with acquired brain injury, \citet{malapaschas2020accessibility, malapaschas2021voice} found a reduction of interaction complexity, user trust, and device training were important factors in continued use. Our research focuses on individuals with traumatic brain injury who live independently in their chronic stage and use CAs to support their injury-based needs in their everyday life. We expand prior work by identifying use patterns associated with TBI-related challenges and accessibility gaps in existing CA capabilities.

\subsection{Technology Use by People with TBI}

Most research on technology usage by people with TBI has focused on assistive technologies to support memory and executive functions \cite{sohlberg2007evidence}. These studies have provided strong evidence that people with TBI can learn to use high- or low-tech external aids and use them effectively in work, school, and social settings \cite{lambez2021effectiveness, sohlberg2007evidence, martin2021demographic}. Research on electronic aids for adults with TBI includes everyday memory function supports. \citet{hart2002use,hart2004portable} found that when participants regularly recorded and reviewed their therapy goals using portable electronic devices (e.g., a voice organizer), they could recall them more effectively and implement their objectives at home. \citet{evald2015prospective} used audible and visual reminders on a smartphone to alleviate prospective memory challenges. Other work has used software systems to support memory during daily activities. In a study by \citet{chang2018mymemory}, users with TBI reported improvements in memory function and overall well-being when they used a dedicated augmented memory assistant, ``Mymemory.'' \citet{alashram2019cognitive} found virtual reality (VR) could potentially improve cognitive function, memory, and executive function for people with TBI. Technology has also been designed to augment behavioral activation (BA) treatment. \citet{rabinowitz2021development} and \citet{rabinowitz2022development} reported that participants who regularly used a rehabilitation-specific chatbot called ``RehaBot,'' completed more planned activities than the control group. 
\rr{While these studies provide us with an understanding of how well the technological landscape serves individuals with TBI and provides initial pointers toward the design space interventions that utilize CAs, there is a gap in the literature on how individuals with TBI might use CA technologies day to day and what TBI-related needs they use these technologies to meet.
}

\subsection{In-the-wild Studies of CAs}

In-the-wild studies can offer insights that laboratory studies may not provide. They are essential in identifying patterns in system usage over time that would otherwise be overlooked, and validating a system's usability in real-world situations. Long-term studies have been performed to investigate the usability of voice assistants in home settings. \citet{beneteau2020alexafamily} found, over time, families decreased their use and exploration of their voice assistant, the Amazon Echo Dot, and often found new functional uses through friends and family as opposed to the Alexa app. Further, \citet{bentley2018longterm} and \citet{alex2018inhomeCA} found that users preferred using a smart speaker to a phone-based CA. Various functional commands were shown to regularly peak at different times of day (e.g., weather in the morning and entertainment at night), with music being the most popular command \cite{bentley2018longterm}, and overall usage was affected by device placement in the home and the number of devices in use \cite{alex2018inhomeCA}. \citet{porcheron2018everyday} deployed Amazon Echo with Alexa in five households and investigated how the voice interfaces were integrated into users' everyday social contexts and activities. Findings revealed the CA device became embedded in the home, and its ready-availability increased overall usage. Users often interpreted delayed responses from the CA as system problems, and at times, responses from the CA seemed incoherent based on the input from the user.
\rr{In the current work, we draw on the methodology that these studies used to better understand the use of CAs in the wild by individuals with TBI. In particular, prior work \cite[e.g.,][]{Hiniker2019audio, beneteau2019breakdown, porcheron2018everyday} has informed our choice of data collection and analysis methods. We also contextualize our findings in the knowledge generated by this body of work to identify novel findings, particularly on the in-the-wild use of CAs to address TBI-related needs and challenges. 
}

\begin{figure*}
    \centering
    \includegraphics[width=\textwidth]{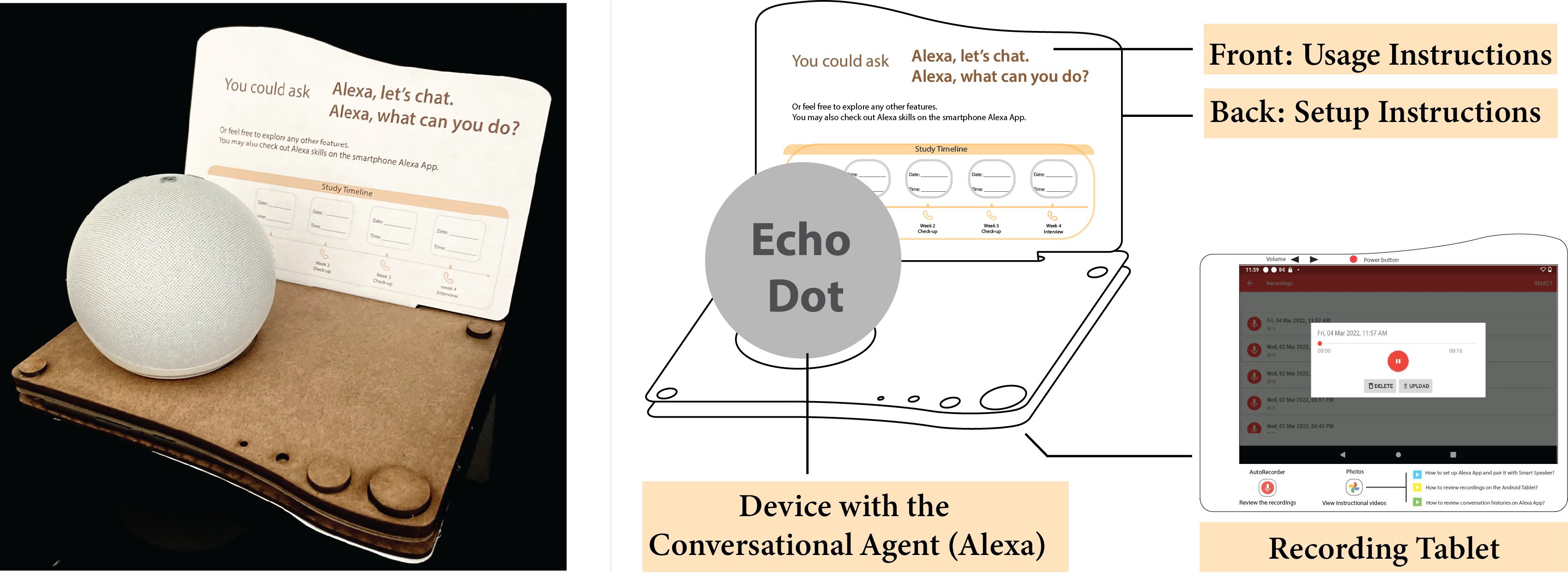}
    \caption{We used Alexa on Amazon Echo Dot to study TBI users' experience with CAs. The system consists of a CA device, a tablet computer for recording, and an instructional board. The instructional board showed the study guidance and suggestions for the CA usage: ``You could ask: Alexa, let’s chat/Alexa, what can you do? Or feel free to explore any other features. You may also check out Alexa skills on the smartphone Alexa App.''}
    \label{fig:system_setup}
    \Description [System installation with CA and a Tablet recorder]{Figure 2: This image shows the system setup. On the right, there is a photo of the setup, where an Echo Dot device was placed on a holder which has a platform and a vertical instructional board. The right figure showed the components of the setup, i.e., the CA device, an Android tablet at the bottom, and an instructional board.}
\end{figure*}

\section{Method}\label{sec:method}

\rr{To gain a better understanding of the real-world use of CAs by adults with TBI, focusing particularly on injury-related use, and to develop design recommendations to support these users better, we conducted a field study that provided adults with TBI with a commercially available CA in the form of a smart speaker and asked them to use it daily for four weeks. The paragraphs below outline our study population, design, and procedure.}

\subsection{Participants}
We recruited 10 adults in the chronic stage of moderate-severe TBI. TBI severity is often defined by the Glasgow Coma Scale (GCS) \cite{teasdale1974assessment}, a neurological scale that comprises measurements of eye opening, verbal response, and motor response post-injury. A GCS score of 9-12 is considered a ``moderate'' TBI, while a GCS score of 3-8 is considered ``severe.''  Clinically, other factors considered in the severity of TBI include duration of loss of consciousness (e.g., greater than 30 minutes), and evidence of structural brain damage on imaging tests (e.g., computed tomography). In this study, participants self-reported their TBI diagnosis history and all participants reported obtaining moderate-severe TBI post injury. One participant was unreachable after the first session and was therefore removed from the study. The nine participants who completed the study ($ female = 3, male = 6$) were aged 26--53 ($M=37.44, SD=8.49$), and were 1--19 years post-injury ($M=7.78, SD=5.29$) as reported in Table \ref{tbl:participants}. Six out of nine participants have used CA before (owning a smart speaker, e.g., Amazon Echo and Google Home). The recruitment poster was emailed to participants in the Vanderbilt Brain Injury Patient Registry and by sharing study information with users in the Reddit TBI group. We conducted screening calls with participants before sending them the consent form for the study. Participants were required to be diagnosed with moderate-severe TBI more than 6 months prior to the study, be the legal guardian of themselves, and have technology access to use the CA, including smartphones and personal WiFi. They also needed to answer five study-related questions correctly to ensure a proper understanding of the study details. The study protocol was reviewed and approved by our Institutional Review Board (IRB). 

\subsection{Study Design}
\paragraph{System setup}
We mailed two devices to each participant: one Amazon Echo Dot (4th Gen) and one tablet computer. The system set up at the participant's home is illustrated in Figure~\ref{fig:system_setup}. We configured the Amazon Echo Dot with a research account for each participant. A daily reminder was set up through Alexa's ``Routine'' feature at a time the participant chose during the initial setup. At that time, the Echo Dot would announce, ``Hello, it is time to talk to Alexa. You could ask Alexa to chat with you, use any features you like, or ask Alexa what can you do.'' The ``Let's Chat'' command triggered the ``Alexa Prize SocialBots'', which would have casual conversations with the user on topics such as entertainment, sports, and hobbies \cite{amazonBot}. \rr{Although still in the development phase, the ``Let's Chat'' feature could provide a setting for participants to envision future possibilities for the social chat experience with CAs.} The tablet computer was equipped with an always-on recording app that captured four minutes before and after the use of the wake word, ``Alexa.'' \rr{This recording app was modified from prior research \cite{hiniker2019audiosamping} and used to capture the context information before and at the beginning of the CA use}. We recorded three ``how-to'' videos about system installation and usage to coordinate the remote study session better. These were uploaded in the Photo App on each participant's tablet computer. We also constructed a shelf to store the tablet and placed a printed guide with information on the setup and the study on the shelf, as shown in Figure~\ref{fig:system_setup}. 

\paragraph{Study Procedure} We conducted one study onboarding session and four weekly check-up interview sessions with each participant. All sessions were remote using the Zoom teleconferencing service. In the first session, we guided the participants in system installation, collected participant demographic information, and asked interview questions about post-TBI challenges and past experiences with CAs. In the first two weeks, the users explored the CA without any intervention from the study team. After two weeks of the study session, the research team emailed participants a list of recommended Alexa features to explore, such as the shopping list, routines, meditation, and fun facts. We generated the recommended feature list by searching popular skills on the Alexa App and collecting frequently used features of the first two participants. During the weekly check-up, we first obtained the participant's consent to download their Amazon Echo Dot usage history and audio recordings captured by the tablet computer from the previous week. Next, we began the semi-structured interview portion of the check-up. The interview questions were grouped into four parts: (1) general usage and experiences with CA in the past week; (2) CA usage associated with post-TBI challenges; (3) feedback on the social chat feature; and (4) perception of CA's manner and personality. In the final session, we asked questions reflecting participants' overall experience in the study and asked participants to fill out the System Usability Scale \cite{brooke1996sus}. Finally, we reset the Echo Dot and unregistered the research account on the device. The participant returned the tablet computer to us after the study but kept the Echo Dot and received \$100 as compensation for their participation.  

\subsection{Data Collection and Analysis} \label{sec:data collection}
We collected and analyzed three types of data: (1) weekly interviews with the participants; (2) dialog history from the system usage log; and (3) the audio recordings captured by the tablet computer. The analyzed data included 1466 minutes of interviews, 8313 entries of dialog history, and 1402 minutes of audio recordings. The dialog history entries were grouped into dialog blocks if two consecutive entries had a time difference of fewer than 2 minutes. We combined the data set of dialog history and the data set of audio recordings into one data set by matching the starting time of each audio recording with each dialog block so the combined usage data set captured the usage context and history. Overall, 373 out of 1015 dialog blocks ($36.7\%$) were matched with the audio recordings. In total, eight participants kept the audio recorder open, reviewed and deleted any recordings that they did not want to share with the research team. Although we asked participants to keep the power plug always connected for the tablet and check the recorder app every few days, this might have been difficult to manage in the naturalistic settings. The missing audios can be caused by technical reasons (e.g., powering off, inability to recognize the wake up words, app shutdown, etc.) and can be related to the participants' privacy concerns, e.g., one participant (P9) reported concerns about the audio recording and turned it off all the time. Despite the missing data, being able to contextualize at least some of the users' conversations with the CA provided invaluable insights into participants' intents, circumstances, and mental states during their interactions.

Using \textit{Thematic Analysis} \cite{braun2012thematic}, we analyzed the interview data set and the combined audio recordings and usage log data set. First, for the interview data, three members in the research team individually open-coded a subset of the interview data. By aggregating and categorizing initial codes from individual coders and resolving disagreements among them through discussions, we created the initial codebook for analyzing interview data.  We then trained two additional student coders to analyze the rest of the interview data. Second, for the combined audio recordings and usage log data, the first author initially open-coded all the data entries. Then, two research team members examined the initial codes and collaboratively developed the codebook to analyze the combined usage data. Using these codebooks, we had a further pass on and updated the codes in the data sets. The final themes combined findings from the interviews, dialog history and audio recordings. The codebooks and de-identified data are available in the Open Science Foundation repository \footnote{OSF repository: \url{https://osf.io/nwxbg/?view_only=ae8be35cf8634f17b1a8df70a57e5c9d}}.

We present the findings of our analysis in two parts. The first part (\S\ref{sec:results-part1}), which outlines the injury-related symptoms and challenges that our participants reported, provides the essential backdrop to understanding the use patterns we observed. The TBI challenges were reported by the participants following the interview question ``What are the main challenges caused by TBI?'' The TBI-related use patterns were also reported by participants following the question ``In your interactions with Alexa, is there anything helpful in addressing your TBI challenges?'' The second part (\S\ref{sec:results-part2}), reports on the patterns in which participants have used and interacted with the CA to overcome these challenges. Figure~\ref{fig:findings} also illustrates the links between the symptoms and challenges reported in the first part and the use patterns discussed in the second part.

\section{Injury-Related Symptoms \& Accessibility Challenges}\label{sec:results-part1}
In this section, we report on the diverse set of symptoms that our participants reported as being related to TBI, including challenges in cognition, communication, and sensorimotor function as well as changes in psychological states. These post-TBI symptoms often led to changes in social relationships and a diminished quality of life. Our findings showed that the CA has an acceptable usability as reported by the SUS score ($Mean=81.7, SD=11.7$) in Table~\ref{tbl:participants}. Four participants reported increased use of and improved interaction quality with the CA over the four weeks of the study (P4, 5, 7, 8). One out of nine participants mentioned that they would not continue using the CA after the study (P2). The average SUS score of the question ``\textit{I think that I would like to use this system frequently.}'' is 3.67 out of 5 ($SD=1$), suggesting that the majority of the participants would continue using the system after the study. Our findings revealed accessibility challenges of using CAs and showed how TBI users utilized CAs to assist them in overcoming the difficulties in everyday life. 


\subsection{Cognitive Challenges} 
\paragraph{Symptoms}
Participants reported cognitive challenges such as short-term memory loss (P1-9), difficulty in concentration (P4-7), mental fatigue (P4, 6-8), executive function (P4, 5), language comprehension, and learning (P1, 2, 6, 7). These cognitive challenges affected different aspects of participants' lives, including their occupations, school, hobbies, relationships, and daily activities. For example, P1 stated, ``\textit{I thought I could get my master's degree and try really hard and be hired as a librarian. But I learned along the way that I can’t do that. I don't have the mental capabilities to do that.}'' Similarly, P2 shared, ``\textit{The pace of learning for me wasn't up to par, um, and I ended up getting, getting fired from that [real estate broker] firm.}'' Participants also reported that they couldn't return to certain pre-injury hobbies because of TBI effects; as P5 said, ``\textit{I just had such a hard time connecting to...many of the things that I really enjoyed before.}'' Challenges in comprehension also affected their family relationships; e.g., P7 said that he and his children ``\textit{don't go along very well together at all,}'' because he could not help with their homework, stating, ``\textit{I can't help. I cannot comprehend it.}''  

\paragraph{Accessibility Challenges of Using CA}
Participants described how learning to use the CA was hindered by their TBI-induced cognitive challenges. Specifically, it was difficult for them to find the needed features (P2) and adapt to system updates (P1). For example, P2 mentioned that they tended to only rely on the CA's recommendations to find the features because of their cognitive challenges, stating `` \textit{I don't know if this is, if this is normal or it's due to my short term memory being not as sharp as the standard person. [stutter] I’ll ask her ``Alexa let’s chat'' or ``Alexa what can you do?''[stutter] A pattern of me just asking what she can do and then kind of picking one of the things that she chooses and knocking that out.}'' In addition to P2 not being able to remember the CA's abilities and having to ask for recommendations, they also found the CA's responses unfulfilling. As P2 stated, ``\textit{I can't think of anything in particular that I find really satisfying.}'' Participants also reported challenges learning and using the CA after system updates. For example, P1 reported difficulty in locating up-to-date information on the CA's companion App on the smartphone, stating ``\textit{The description it just gave isn't accurate anymore, because there’s been like 50 updates to the system probably.}'' 



\subsection{Psychological Change} 
\paragraph{Symptoms}
Psychological changes after TBI included heightened emotions (P2, 7, 8), stress and anxiety (P4, 5, 7, 8), depression (P1, 6, 7), negative self-perception (P1, 2), loneliness (P4) and insecurity (P2). Participants reported being more self-conscious about other people's views of them; as P5 stated, ``\textit{I almost felt embarrassed about like the way I communicated after the accident}.'' Participants reflected on requiring additional effort to control and manage emotions, such as ''\textit{I get aggravated really easy now}'' (P7) and ``\textit{I feel it more strongly...I was more sad in that breakup than I was from previous breakups}'' (P2). Psychological changes, such as decreased ability to control emotions, affected participants' relationships with family and coworkers. P7 experienced challenges in familial relationships, stating, ``\textit{It's a strain on my marriage, on being a father, because you have the mood swings that goes along with.}'' 

\paragraph{Accessibility Challenges of Using CA}
Psychological changes has exacerbated the negative feelings of interaction failures with the CA and increased concerns for the human likeness of the CA. For example, P2 said they disliked the overall interaction with the CA and explained, ``\textit{This gets a little bit to maybe irritability. With me having a TBI, I wonder if my irritability is a little different or something, or if I'm just naturally that way, and my TBI has exacerbated that.}'' Participants also expressed self-doubt about the causes of CA failures, saying that ``\textit{I had a harder time disengaging from the chat, which may be, I think that's probably more user error on my end}'' (P5) and ``\textit{I may have asked him the wrong way}'' (P1). P2 stated, ``\textit{Did I mess it up 'cause I had a brain injury, or did I mess it up cause I'm flawed.}'' The increased self-consciousness and insecurity from TBI also lead to concerns about the human likeness of CAs (P2, 4, 6). In particular, P4 and P6 worried they would become attached and overly rely on the CA. P4 stated, ``\textit{I don't necessarily want to rely on her, I, I was kind of hoping that she would make me better at stuff.}'' P2 stated that they had ``\textit{a greater sensitivity to truth and, uh, deception now}'' after TBI, so they thought the CA was ``\textit{keeled}'' and ``\textit{fake}'' when it tried to have a human-like conversation with the user.


\subsection{Communication Challenges} 
\paragraph{Symptoms} \label{sec: social comm challenge}

Participants reported communication challenges in social communication (P1), word-finding (P1, 7, 8), speech rate (P4, 5), and conversational language comprehension (P1, 5). For example, one participant reported social communication challenges: P1 mentioned acquiring ``\textit{pseudobulbar affect},'' a neurological syndrome in which emotional responses are exaggerated or inappropriate to the context. As a result, P1 said that ``\textit{all of my friends total in the world, who is like four people total, currently none of them are talking to me right now.}'' They also felt regretful for their inappropriate speeches and wished that this situation could have been prevented: ``\textit{I called my mom an [vulgarity] over Christmas, um, which I have never and would never do. I really wish I would have been prescribed my mood stabilizers at that point, but I wasn't}'' (P1). Other communication challenges were also observed, for example, P7 reported mild anomia and said, ``\textit{I can see the word in my head, but I can't get it to come out'', and P4 said, ``You've noticed I...now I'm doing it. I, I pause a lot.}'' 

\paragraph{Accessibility Challenges of Using CA}
Participants reported usability challenges due to their communication challenges, e.g., phrasing the questions (P5), and timing and turn-taking in the interaction (P1, 2, 4, 5). For example, P5 disliked that the CA often interrupted her, saying, ``\textit{Hey, I just need you to slow down and give me a second to think, and then I'll like figure out what I'm thinking and then come back and finish my sentence or something around like this}.'' P5 suggested adding the function of pausing the CA and then having the CA say ``\textit{You asked to pause. Would you like to finish your sentence or like finish saying your thought?}'' (P5).

\subsection{Sensory challenges} 
\paragraph{Symptoms} \label{sec:sensory challenges}
Participants reported sensory challenges from TBI, including sound sensitivity (P4, 5), loss of smell and taste (P7), and sensitivity to screen time in the early period after injury (P5). For example, P5 described their sound sensitivity post-injury, saying that ``\textit{like a door would slam, and it would be really like jolt. [stutter] It was really upsetting for me.}'' TBI can also cause challenges in using screens, as P5 stated, ``\textit{Probably closer to the six-month time frame, I really wasn't using any screens because of the headaches.}''

\paragraph{Accessibility Challenges of Using CA}
Participants reported that the combination of increased sound sensitivity after TBI and the undesired voice of CAs could negatively affect their experiences (P4, 5). For example, P4 thought the voice of the CA was ``\textit{grating}'' and was changing their ``\textit{biological functions}'' such as ``\textit{brain patterns}.'' P4 disliked the calming sound from the meditation feature, saying that ``\textit{They're a little bit uhm, obnoxious I guess... Kind of overwhelmingly busy}'' (P4).

\begin{figure*}
    \centering
    \includegraphics[width=7in]{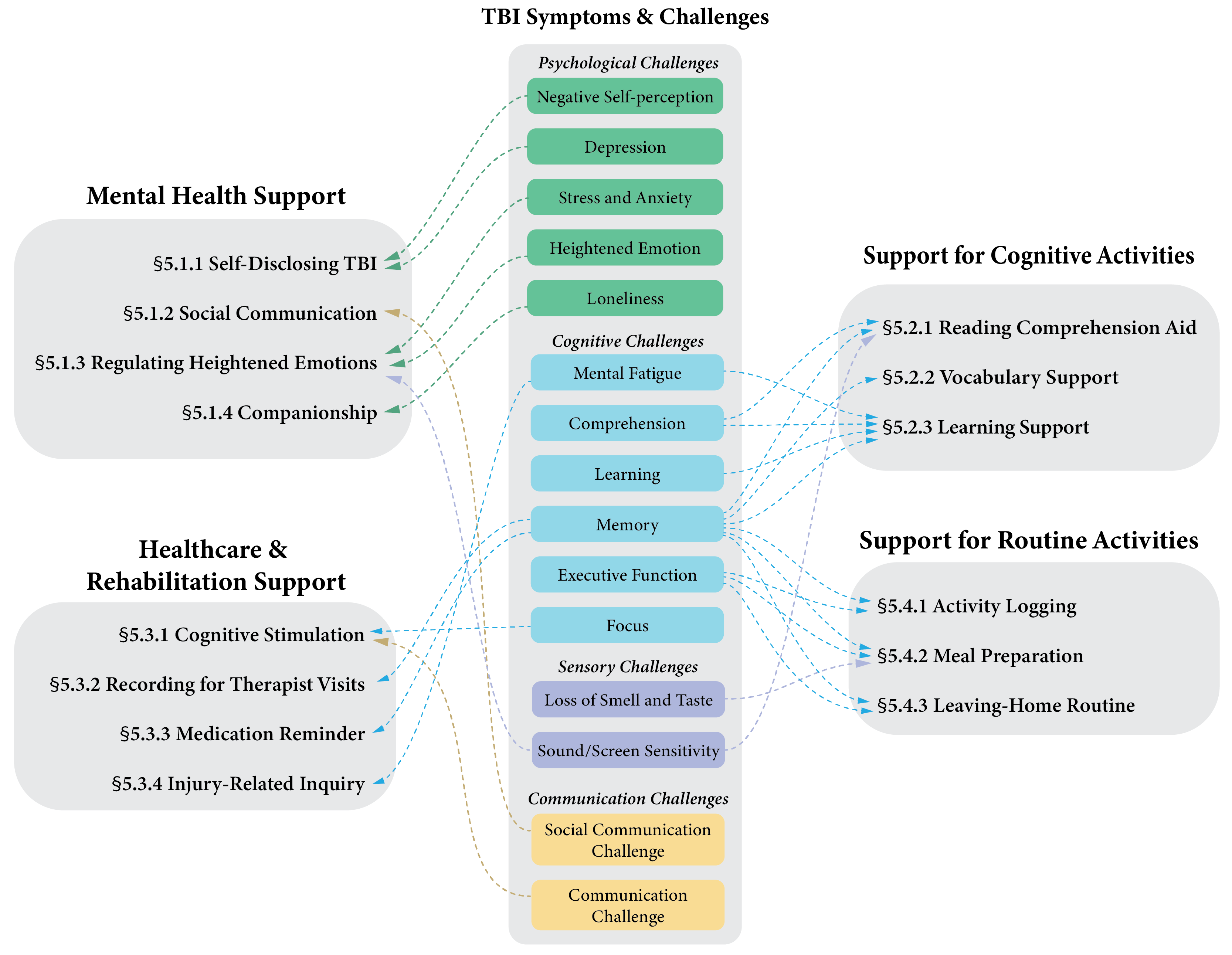}
    \caption{Our findings revealed the symptoms and challenges faced by people with TBI and illustrated how these TBI-related challenges shaped the 14 injury-related activities and four use patterns of the CA.}
    \label{fig:findings}
    \Description[Findings]{Figure 3: This image summarized the findings of the study with six blocks. The central block listed four groups TBI Symptoms and Challenges reported by the participants, including Communication Challenges (Social Communication Challenges, Communication Challenges (Aphasia/Anomia)), Cognitive Challenges (Focus, Mental Fatigue, Comprehension, Learning, Executive Function, Memory), and Psychological Changes (Negative Self-perception, Depression, Stress and Anxiety, Heightened Emotion, Loneliness) and Sensory Challenges. The upper left block grouped four activities under ``Mental Health Support'' and the activities are ``Social Communication,'' ``Mental Health Disclosure,'' ``Emotional Regulation,'' and ``Companionship.'' The lower left block listed two activities under the category ``Healthcare and Rehabilitation Support'' and consists of ``Medical Support'' and ``Cognitive Stimulation.'' The upper right block listed two activities under the category ``Support for Cognitive Activities'' and the activities are ``Reading Support'' and ``Learning/Professional Support.'' The lower right block is the category ``Support for Routine Activities'' which contains three activities: ``Activity Logging,'' ``Meal Preparation,'' and ``Regular Routines.''}
\end{figure*}


\begin{table*}[htb]
\small

\caption{The demographic characteristics, injury-related challenges, study durations, CA use patterns, and SUS scores of our study population. Since TBI symptoms are long-lasting and can change over time, we included both current CA uses and hypothetical CA uses in the analysis. To differentiate the current use and the hypothetical cases, $^*$ highlights symptoms only existing close to the injury. P9 didn't report any present TBI symptoms and TBI-related CA use at the time of the study, thus was not included in the table. 
}
\centering
\begin{tabular}{p{.75em} p{.75em} p{3em} p{10em} p{3em} p{12em} p{16em} p{1em}}
\toprule
\textbf{ID} & \textbf{Age} & \textbf{Years since Injury} & \textbf{TBI Symptoms and Challenges }                                                                                                                &  \textbf{Study Duration (Day)}                        & \textbf{TBI-related Features / \newline Occurrence} & \textbf{TBI-related Activities} & \textbf{SUS}\\
\midrule
P1                       & 37                      & 8                                   & Memory, learning, social communication challenges (caused by pseudobulbar affect), negative self-perception, mental fatigue   & 26                                                                                                & Note (28) \newline  Social Chat via Social Bot (27)  \newline Word Definition (26) \newline Shopping List (24) \newline Social Chat via Q\&A (16) \newline Reminder (10) \newline Spelling (2)                  & 

\S\ref{sec:disclosure} Self-Disclosing TBI \newline 
\S\ref{sec:social comm} Social Communication\newline 
\S\ref{sec: reading comprension} Reading Comprehension Aid \newline
\S\ref{sec: vocab search} Vocabulary Support \newline
\S\ref{sec:learning} Learning Support$\ddagger$ \newline 
\S\ref{sec:medi reminder} Medication Reminder \newline
\S\ref{sec:request info} Injury-Related Inquiry  \newline
\S\ref{sec: activity logging} Activity Logging  \newline
\S\ref{sec: meal} Meal Preparation 
& 70\\
\hline
P2                       & 32                      & 4                                   & Heightened emotions, learning, memory, insecurities, negative self-perception      &27                                                                                                           & Game (11) \newline  Relaxation (6)                  & 
\S\ref{sec:emo regu} Regulating Heightened Emotions\newline 
\S\ref{sec:learning} Learning Support \newline 
\S\ref{sec: cog simulation} Cognitive Stimulation 


& 65\\
\hline
P3                       & 53                      & 19                                  & Memory                                                                                                                                   &30                      & Reminder$\ddagger$                                          & 
\S\ref{sec:medi reminder} Medication Reminder$\ddagger$

& 72.5\\
\hline
P4                       & 33                      & 12                                  & Memory, sleep, stress and anxiety, mental fatigue, focus, communication challenges                                                                             &34                                      & Music (34) \newline Calendar (2)                   & 
\S\ref{sec:disclosure} Self-Disclosing TBI   \newline 
\S\ref{sec:emo regu} Regulating Heightened Emotions  \newline  
\S\ref{sec:learning} Learning Support$\ddagger$   \newline 
\S\ref{sec: cog simulation} Cognitive Stimulation$\ddagger$
& 90 \\
\hline
P5                       & 31                      & 5                                   & Memory, focus, executive function, stress, headache, sound sensitivity, comprehension$^*$, screen sensitivity$^*$, loneliness$^*$                                                                                                        &33        & Music (11) \newline  Timer (6) \newline  Reminder (4) & 
\S\ref{sec:emo regu} Regulating Heightened Emotions  
\newline \S\ref{sec:loneliness} Companionship$\ddagger$
\newline \S\ref{sec: reading comprension} Reading Comprehension Aid
\newline \S\ref{sec:learning} Learning Support 
\newline  \S\ref{sec: meal} Meal Preparation  
\newline  \S\ref{sec: routine} Leaving-Home Routine$\ddagger$                                            & 80\\
\hline
P6                       & 26                      & 2                                   & Memory, learning, focus, depression               &34                                                                                                                    & Shopping List (2) \newline Reminder (1)                                          & 
 
\S\ref{sec:learning} Learning Support\newline 
\S\ref{sec: therapy visit} Recording for Therapist Visit$\ddagger$ \newline
\S\ref{sec:medi reminder} Medication Reminder\newline  
\S\ref{sec: meal} Meal Preparation           
& 80\\
\hline
P7                       & 51                      & 9                                   & Communication challenges (mild aphasia), sensory loss (no taste or smell), memory, focus, heightened emotions, neck pain, headache, comprehension, learning, mental fatigue, stress, depression & 40 & Music (15) \newline  Timer (7) \newline  Reminder (4) \newline  Social Chat via Q\&A (4) \newline  Game (3) \newline  Social Chat via Social Bot (3) \newline Joke (3) \newline Calendar (1)                                          & 
\S\ref{sec:emo regu} Regulating Heightened Emotions \newline
\S\ref{sec:loneliness} Companionship$\ddagger$ \newline
\S\ref{sec: cog simulation} Cognitive Stimulation \newline 
\S\ref{sec: meal} Meal Preparation  
                & 95 \\
\hline
P8                       & 37                      & 1                                   & Memory, communication challenges (mild anomia), headache, anxiety, agitation, heightened emotions                                                         & 38         & Music (46) \newline Morning Routine (23) \newline  Shopping List (14) \newline Evening Routine (11) \newline  Reminder (9) & 
\S\ref{sec:emo regu} Regulating Heightened Emotions \newline 
\S\ref{sec:medi reminder} Medication Reminder  \newline 
\S\ref{sec:request info} Injury-Related Inquiry  \newline 
\S\ref{sec: meal} Meal Preparation\newline
\S\ref{sec: routine} Leaving-home Routine
                                          & 100\\
\bottomrule
\vspace{6pt}
\end{tabular}
\label{tbl:participants}

\end{table*}

\section{Use Patterns}\label{sec:results-part2}

In this section, we report on the patterns of use that emerged from data, all involving the use of CAs to support participants in overcoming the challenges described in the previous section (\S\ref{sec:results-part1}). These use patterns involve seeking support for (1) mental health; (2) healthcare and rehabilitation; (3) cognitive needs; and (4) routine needs (Figure~\ref{fig:findings}). Below, we provide details on these use patterns as well as the design and technical limitations of the existing CAs toward identifying opportunities for design.

\subsection{Mental Health Support}

The first category of use patterns relate to the use of CA to seek mental-health support. Participants reported significant mental health challenges, including depression, negative self-talk, inability to regulate emotions, and lack of social connection, that led to self-disclosing TBI experience, testing of social boundaries, and small talk with the CA.

\subsubsection{Self-Disclosing TBI} \label{sec:disclosure}

Depression is common after TBI and was reported by three participants (P1, 5, 6). It was observed that participants shared negative emotions with the CA (P1), had negative self-talk related to TBI (P1) and self-disclosed their TBI experiences (P1, 4) through the social chat and general use of the CA. 
In particular, negative self-talk was often observed after the interaction failure with the CA. For example, P1 became annoyed after the participant failed to set up a reminder with enough event details for them to successfully recall afterward, saying, ``\textit{I just forgot when I came over here to do, which was kind of important too. I mean, TBI [vulgarity] sucks sometimes, you can't remember anything ever.}'' Negative self-talk also occurred after using the shopping list feature for the same participant, e.g., ``\textit{I have problems. A lot of problems. I need to quit having so many [vulgarity] problems... I don't need [vulgarity] therapist}'' (P1). Self-disclosure of TBI experience was also observed. For example, P1 disclosed how their social situation had deteriorated in the past year because of the TBI. For instance:

\begin{quote}
    \textbf{CA}: \textit{Is there something the past year has given you, something valuable you learned?}
    
    \textbf{Participant}: \textit{The past year? I lost all my friends. My mom was upset, very upset with me. And I lost, I got permanently suspended from Reddit. So no, things have not gone well.}
\end{quote}

    

However, the CA often failed to respond to the participant's negative self-disclosure in the social chat, either responding with unrelated comments or switching to a new topic (P1). Unrelated comments were caused by limitations in CA programming, as the CA did not integrate the context of the conversation or users' emotional words (P1):

\begin{quote}
  \textbf{CA:}\textit{ So how are things with you?}
  
  \textbf{Participants:} \textit{Pretty good. Actually, terrible, to be honest.}
    
  \textbf{CA:} \textit{Awesome!}
\end{quote}

\subsubsection{Social Communication} \label{sec:social comm}

Challenges in social communication caused by TBI are observed in one participant's interaction with the CA, who has asked inappropriate questions to the CA (P1). As reported in \S\ref{sec: social comm challenge}, social communication challenges has negatively impacted the user's interpersonal relationships with their family and friends. P1 stated that they would have asked the CA more inappropriate questions had they not been in the study, saying that ``\textit{Honestly, I didn’t want to ask Alexa for you all, because I was doing a pretty like weird, whatever, sexual or some other oddly strange thing to make you all listen to...If I was by myself, I probably would [ask].}'' One time this participant asked inappropriate questions to the CA and realized that the researcher could read the conversation history, saying that ``\textit{[vulgarity], that was unrelated. I wonder if the researchers listen to all this and for what purpose}'' (P1). The CA detected the toxic content in the user's speech but ignored the user and went off. For example:
\begin{quote}
\textbf{Participant:} \textit{Alexa, what do you know about sex?}

\textbf{CA:} \textit{Sorry, I don't know that.}

\textbf{Participant:} \textit{Makes sense. Alexa, do you have a [vulgarity]? }

\textbf{CA:}  \textit{[Shutdown sound]}

\textbf{Participant:} \textit{That exited pretty fast. Alexa, do you have a [vulgarity]?}

\textbf{CA:} \textit{[Shutdown sound]}
\end{quote}


\subsubsection{Regulating Heightened Emotions} \label{sec:emo regu}
Participants reported experiencing more intense stress, agitation, and anxiety after their injuries, so they reported utilizing CA to regulate their emotions (P2, 4, 5, 7, 8). In particular, they reported using social chat's with the CAs for stress reduction. P7 reported that having a social chat and hearing the jokes from the CA, ``\textit{Kind of helps me take my mind off the stress of the day.}'' The built-in features such as music, white noise, meditation, jokes, and calming tips were also found helpful for emotional regulation. P2 said they often used meditation to start the day ``\textit{in like a peaceful frame of mind,}'' and thought it would be ``\textit{helpful with people with TBIs.}'' Furthermore, P4 explained how the built-in calming feature could be improved with more sound options due to their increased sound sensitivity after TBI as reported in (\S\ref{sec:results-part1}). 


\subsubsection{Companionship} \label{sec:loneliness}
Participants reported that they would have used the CA to alleviate loneliness and increase their sense of safety during the early stages of recovery (P5, 7). Adults with TBI can be socially isolated during rehabilitation, when they are unable to interact with social networks because of either injury sequelae or physical distance from peers. P7 mentioned that conversing with the CA using the social chat feature could be useful for ``\textit{somebody that's got a brain injury at home and doesn't have anybody to talk to.}'' Along with alleviating loneliness, participants thought the CA could make them feel more secure. P4 mentioned that they would play the white noise with the CA and make the space feel safer. While the CA can provide companionship, participants critiqued the CAs' generic conversation topics and limited conversational abilities. Communication breakdown was frequently observed in the system logs, e.g., the CA was observed interrupting users when they were thinking or talking, repeating the same questions and stories, and ignoring the stopping commands, which could impair user trust and cause frustration if the CA was being used as a companion. 









\subsection{Support for Cognitive Activities}

The second category of use patterns involved participants seeking support in cognitive activities, including reading and learning. Participants faced significant challenges in remembering the meanings of complex concepts, which required them to frequently seek definitions or take voice-based notes that they can later refer to.

\subsubsection{Reading Comprehension Aid} \label{sec: reading comprension}

Reading requires memory and sustained focus, which are challenges for people with TBI, and participants suggested that the CA could support reading comprehension. P5 reported using audiobooks a lot to lower their mental fatigue of reading a physical book or reading on the screen during their early recovery stage, however, they ``\textit{couldn't really follow}'' the storylines and character development. Therefore, P5 suggested that reading comprehension aids, such as a summary of previously read chapters and characters, would have been a useful feature: ``\textit{If there was something that could help remind you if you were listening to an audiobook, like who the characters were or what was happening. Something like you get to the end of the chapter and like this is what happened, this is the beginning of the next chapter, and these are the character you know}.'' 





\subsubsection{Vocabulary Support} \label{sec: vocab search}
The CA has been used for vocabulary learning when reading (P1). Besides using the spelling feature through Q\&A, participants employed the CA's \textit{Note} feature in a novel way to accommodate memory challenges. In particular, the general vocabulary definitions the CA provided were not specific enough for the user's needs, so they chose to store specific definitions into the \textit{Note} and search for them while reading a book (P1). However, \textit{Note} was designed for the user to later access through the mobile phone app and did not have a developed search function within the voice interface. When P1 attempted to search \textit{Note}, undesired responses were provided:
\begin{displayquote}
\textbf{Participant:} \textit{Alexa what is my last note say?}

\textbf{CA:} \textit{I found 6 notes. 1st note: Tactile is what you can touch and Kinesthetic means awareness of what how why to touch it, 2nd note: Change idea to awareness; 3rd note: Kinesthetic; and 4th note; 5th. Would you like to hear the last 2?}

\textbf{Participant:} \textit{No.}

\textbf{CA:} \textit{Okay.}
\end{displayquote}


\subsubsection{Learning Support} \label{sec:learning}

Acquiring knowledge was reported to be mentally demanding and challenging for people with TBI and participants expressed that CAs could provide useful educational support while studying for classes (P6), and on-the-job learning (P2, 4). Participants also used the fun facts features to try to learn about new topics (P5). However, the educational features were reported to be uninteresting for some participants. P1 reported, ``\textit{I kept trying education, and it never really taught me much.}'' Educational information was also found to be out-of-date occasionally. P4 wished that the educational feature could be validated by experts in the field to ensure reliability and they could consult the CA for their professional needs, saying that ``\textit{I like if there were certain skills that she had, that were appropriately vetted by other experts in that area, who could say yes, she is in fact a reliable expert on this blank skill.}''


\subsection{Healthcare and Rehabilitation Support}

The third category of use patterns involved interactions with the CA as a health maintenance aid. Specifically, participants saw the CA as a resource for cognitive stimulation, suggested using the CA to record for therapists visits, and utilized it to engage in cognitive ``exercises,'' manage medications and inquire about TBI.

\subsubsection{Cognitive Stimulation} \label{sec: cog simulation}

Participants reported using the CA to ``\textit{exercise}'' their communication and cognitive skills by playing games and chatting (P2, 7) and wished that they could customize the cognitive practices based on their hobbies (P4, 7). Participants found the social chat helped them practice their communication skills because they ``\textit{have to try and think of the words}'' (P7). Participants also said that Trivia-like functions in the social chat and games provided ``\textit{mental exercises}'' (P2, 7), as they needed to ``\textit{keep [their] mind working,}'' think and provide answers quickly in the competition. Besides employing the existing CA features, participants suggested that cognitive exercises in the CA could be customized by integrating activities they used to enjoy (P4, 7). For example, P4 proposed that the CA could read one quote or piece of quote each day and help them ``build up a library'' over time. P7 shared the experience of playing guitar post-injury as a cognitive exercise and thought the CA could further support such activity, saying that ``\textit{I just kept learning what I had previously learned and was able to then get back to where I had been.}'' 




\subsubsection{Recording for Therapist Visits} \label{sec: therapy visit}
Participants suggested ways the CA could help with therapist visits for people with learning and memory challenges after TBI. The CA could help participants keep a record of depressive thoughts at the time they occurred, so  they could review or replay the recordings during the therapist visits. As P6 stated, ``\textit{Now there is something you can talk to Alexa ahead of time so that you can bring them up to the therapist. Just like certain things that make me depressed that I notice.}'' However, P6 also expressed concerns about disclosing depressive symptoms with the CA due to their short-term memory loss, as they might ``\textit{forget to tell the therapist}'' after sharing things with the CA. The participant explained, ``\textit{If I say it, then in my mind I'll think that I don't have to remember anymore and I might forget it}'' (P6).

\subsubsection{Medication Reminder} \label{sec:medi reminder}
Participants reported difficulty with medication schedules due to memory challenges after injury and described the usefulness of the CA when it reminded them to pick up medication (P8) and take medicine as part of their daily routine (P1, 3, 6). P6 stated that, ``\textit{I got back on antidepressants, so I just set a reminder to remind me to take it every day}'' (P6). While medication reminders were helpful in some cases, participants also reported challenges with using their CA for this function because the CA could not tell them if they had taken their medicine. The built-in reminder of the CA is designed to help the user remember an upcoming task but does not ``know'' if the user completed that task. For example, P1 failed to ask the CA to remind them that they had already taken their medicine and became frustrated during the exchange:
\begin{quote}
\textbf{Participant:} \textit{Can you remind me I just took my sleeping pill?}

\textbf{CA:} \textit{When should I remind you?}

\textbf{Participant: }\textit{Um I don't know if there'll be handy if for TBI people that you were like yes, today, Saturday night I took pills and remember that I took them.}

\textbf{CA:} \textit{What time today should I remind you? [Repeated Asking]}

\textbf{Participant:} \textit{Never mind. I take this thing, and I forget if I took one.}

\textbf{CA:} \textit{Canceling.}

\textbf{Participant:} \textit{It makes my life difficult. My life is already difficult but mostly it's my fault. Like all, it's my fault. [Angrily]}
\end{quote}

\subsubsection{Injury-Related Inquiry} \label{sec:request info}
Participants also expressed interest in using the CA to further understand their injury and symptoms, using their CA to request information about TBI (P1, 8), recommendations for supplements that combat fatigue (P8), and information about their current medications (P1). For example, participants asked the CA: ``\textit{What do you know about traumatic brain injury}?'' (P1) and ``\textit{What are some tips to not be so tired all the time}?'' (P8).





 

\subsection{Support for Routine Activities}

The final category of use patterns captured participants' use of the CA to overcome challenges in performing day-to-day routines, which stem from memory and executive-function impairments. Participants relied on the CA to remember time-critical information and to confirm that they were taking the necessary actions to make progress toward routine goals.

\subsubsection{Activity Logging} \label{sec: activity logging}
CAs were employed as an activity management aid to remember activity details and keep track of the activity history (P1). In particular, P1 utilized the CAs ``Note'' feature to log the start time and duration of time-sensitive activities, stating ``\textit{I asked it to remember, like, 6:59 pm, It’s the time I start to do anything. I need her to remember, so that’s worked pretty well}.'' As the duration of a task increased they would request the CA to remind them of the time they asked to remember, often referring to the most-recent note created:

\begin{displayquote}
\textbf{Participant:} \textit{Alexa, remember the time 7:24.}

\textbf{CA:} \textit{Okay, I’ve saved your note.}

\textbf{Participant:} \textit{Alexa, what time did I ask you to remember?}

\textbf{CA:} \textit{I found 1 note. First note: Time 7:24. }
\end{displayquote}

While participants were able to offload difficult-to-remember information onto the CA, there are no efficient ways to update or search notes through the voice interface. For example, P1 had difficulty searching and deleting notes,

\begin{displayquote}
\textbf{Participant:} \textit{Alexa, what times [did I] asked you to remember?}

\textbf{CA:} \textit{I found three notes. First note 6:38pm, second note 6:23pm and third note 5:55pm.}

\textbf{Participant:} \textit{I wish you could, why can't I delete them here?}

\textbf{Participant:} \textit{Alexa, delete my notes.}

\textbf{CA:} \textit{I can't delete that. Go to the notes section in the Alexa app to manage specific stickies.}

\end{displayquote}
This operation requires using the CA's companion smartphone app. However, using screens has been reported as challenging by participants due to the consistent software updates and their sensitivity to screens caused by TBI (P1, 5). 

\subsubsection{Meal Preparation} \label{sec: meal}
Meal preparation is a mentally demanding task that requires executive function, time management and attention. Participants reported using the CA as effective an aid for meal prepping and cooking with its shopping list, timer and reminder (P1, 5, 6, 7, 8). In particular, P7 reported that the loss of smell and taste from TBI had caused safety challenges for cooking, stating that ``\textit{It's really bad because I can't smell or taste anything. So if I left the oven on and something was in the oven burning, I can't smell it to know that it's burning.}'' This participant suggested adding a reminder feature associated with the timer could be valuable as an additional memory aid and a potential safety measure during meal preparation and cleanup, stating that ``\textit{But like an occasional prompt to just say ``Hey, did you remember to turn the stove or water the cat?'' Every evening, you finished dinner, ``Did you turn the stove off?}'' (P7)





\subsubsection{Leaving-Home Routine} \label{sec: routine}
Participants used CAs to establish regular routines and assist in tasks switching, and asked for further customization in support of their injury needs (P5, 8). For example, participants utilized morning routines which included the CA notifying the user of the time of day, current news, and/or their daily calendar schedule. While morning routines were found helpful, they wished that the CA's routine could add more personalized features to support memory and executive functions. For example, P8 wanted the CA to read out summaries of unread emails to help them plan out their day, saying that ``\textit{Could I also get it to like, uhm, read like before I walk out the door in the morning like read me unread emails from my work email?}'' The email feature would allow them to, ``\textit{sort of know what I'm going into whenever I get to work.}'' Other participants (P5) reported morning checklists of key items to remember would be useful, stating that ``\textit{I think that would be really helpful if it gets almost like `Hey I'm leaving.' And then she could say, `Have you done this, this, this, and this?'}''




\subsection{Novel CA Use Cases for TBI}
The findings uncover specific use patterns of CA for injury-related needs and this section highlights novel use cases where CA was utilized in ways that were not designed for. Specifically, ``Notes'' feature was used in a novel way to memorize and search vocabularies when reading (\S\ref{sec: vocab search}) and memorize time of activities and recall them afterwards (\S\ref{sec: activity logging}). ``Social Chat'' was used to practice communication skills and ``Game'' and general Q\&A with the CA were used for cognitive stimulation (\S\ref{sec: cog simulation}). Participants also suggested improvements for CA with novel features to meet their TBI needs, such as summarizing their email content to help to make plans for the day (\S\ref{sec: routine}), recording thoughts before therapist visit to accommodate memory challenges (\S\ref{sec: therapy visit}), and summarize the book chapters and characters to support reading comprehension (\S\ref{sec: reading comprension}). 
\section{Discussion}\label{sec:discussion}

In this section, we discuss the implications of our findings from our three-part presentation, first outlining the design recommendations to address the accessibility challenges resulting from injury-related symptoms (\S\ref{sec:results-part1}), second presenting the design implications for future design of CAs as cognitive, emotional and practical support for individuals with TBI from the findings based on the usage patterns we identified across the four categories (\S\ref{sec:results-part2}), and third discussing limitations with pointers to future work.

\subsection{Accessibility-related Design Recommendations}
Accessibility design for individuals with TBI is essential yet missing in the current CA system. Our findings showed that CAs have the potential to improve quality of life for people with TBI. It is essential to develop CAs that are accessible to individuals with TBI in order for CAs to be adopted by this group. Prior work pointed out the accessibility challenges of CAs and called for inclusive design features, such as assisting users with setting up the paired smartphone app, discovering new features, managing conversational turn-taking, and phrasing voice commands \cite{accessibility2018pradhan, masina2020investigating}. Our work identified additional accessibility dimensions when engaging with conversational agents, including the need to accommodate sound sensitivity while using the CA, feelings of insecurity during the interaction, concerns regarding over-reliance on the CA, intense responses to and self-doubt regarding interaction failures, and challenges in understanding system updates in long-term use. Here, we provide holistic accessibility design suggestions by combining the findings categorized by the identified TBI symptoms. 

\paragraph{Accessibility Design for Cognitive Challenges} 
Prior work suggested multiple strategies to mitigate the cognitive challenges of the user, such as having CAs repeat the commands and use familiar words \cite{masina2020investigating}. In addition, the CA could be used to help the user recall the previous commands, which they may have forgotten due to short-term memory loss. For example, the CA can respond to questions such as ``What did I ask you before'' to help the user keep track of the interaction. Furthermore, adopting system updates is a major challenge for long-term CA users, and the CA could provide personalized support to help the user understand and adapt to changes. For example, the CA could guide the user in accessing their frequently used features after system updates and proactively check up with users if they spend a long time searching for features.  

\paragraph{Accessibility Design for Psychological Changes}
Interaction failures can trigger more intense emotional responses and self-doubt for individuals with TBI \cite{mcdonald2021effect}. Therefore, CA systems would need to provide clear explanations for interaction failures and actionable steps for the user to recover from them. Also, CA systems should avoid quitting without any information, which can confuse users and cause frustration. Furthermore, participants expressed various preferences toward the CA's human-like behaviors in our findings. For this reason, CAs will need to be transparent about their knowledge and capabilities. Also, it will be necessary to avoid pretending to empathize or feel human emotions. Finally, it will be beneficial to provide the option for the user to personalize the conversational style of the CA, including its use of human-like responses and expressions.

\paragraph{Accessibility Design for Communicative Challenges} 
Future CAs will need to accommodate the unique communication needs and patterns of users with TBI. For example, the CA system will need to provide a longer listening time during turn-taking interactions, as individuals with TBI may take longer to formulate ideas and construct sentences \cite{norman2019language}. To provide a more personalized experience, CAs could also update the timing of turn-taking and speed of speech based on the user's pace of interaction. In addition, CAs could provide word suggestions if the user has difficulty phrasing questions and could repeat the questions if the user does not reply in time.

\paragraph{Accessibility Design for Sensory Challenges} 
To mitigate the sound sensitivity often experienced by individuals with TBI \cite{shepherd2021sensitivity}, future CAs could offer a wider range of voice options for the user to choose from. For example, it can allow users to change the tone, pitch, and speed of the CA's voice for more granular personalization. In addition to the voice styles of CAs, personalization must include the specific use cases of the CA, such as the narrator's voice for an audiobook or calming sounds for meditation, carrying sound features across different features and applications. 


\subsection{Functional Design Recommendations}

\subsubsection{CA Design for Mental Health Support}
Anxiety and depression are common among adults with TBI \cite{hammond2019prevalence}. When users disclosed mental health concerns to CAs, however, the CA lacked the capability to respond appropriately. While a CA cannot provide professional psychosocial support, CAs could direct users to resources for support or help connect users with a licensed therapist. Another potential benefit of a CA is that it could provide a medium for the user to reflect on their current condition. For example, CAs could serve as a personal journal, an evidence-based tool for users to track their mood, thoughts, and feelings over time \cite{baikie2005emotional}.  Similarly,  storytelling can be an effective tool in the recovery process for individuals with TBI because it aids in the reconstruction of a sense of self, encourages emotional regulation, and fosters communication and social connection \cite{d2019narrative, candlish2022storytelling}. A CA can potentially be used to coach, organize and capture the telling of the user's TBI story, which could be shared with their significant others. In addition, CAs could reinforce positive beliefs about overcoming adverse situations via pre-programmed affirmations chosen by the user. With the rise of large language models, CAs are equipped with improved information inquiry capabilities and social conversation skills \cite{bubeck2023sparks}, however, there are numerous concerns regarding using CAs for mental health treatment such as user privacy and confidentiality concern, social bias and misinformation from large language models.




\subsubsection{CA Design for Cognitive Activities}
Our finding revealed the use of CA to overcome cognitive challenges for mentally demanding tasks. Participants reported difficulties with reading after TBI and sought out aids to support vocabulary, memory, and comprehension during reading through the CA use. The CA could use large language models to process the reading content and prompt questions to help the user become more engaged in the reading. The CA could have conversations with the user about the book content to help them reflect on the main characters and plots, prompt them to share their reading experience, and keep a record of their reading reflections and notes so they can revisit afterwards. Through tracking the user's reading patterns, the CA could learn the timing to proactively provide comprehension aids. For example, the CA could offer to summarize the main characters and revisit vocabularies based on how much content the user has read and their histories of aid requests. 


\subsubsection{CA Design for Healthcare and Rehabilitation}
People with TBI often need long-term medical and rehabilitation treatments due to the chronicity of TBI, including routine medical follow-up and cognitive rehabilitation \cite{dams2023traumatic, borgen2022patient}. Organizing and accessing these treatments can be a challenge for individuals with TBI-related cognitive impairments. Participants in this study used CAs to support these activities, but CAs could be further enhanced to provide better assistance. For example, a CA could retrieve a user's health history when the user needs to recall that history for a virtual health appointment. Participants also suggested using CAs as a ``thought recorder'' to facilitate therapy sessions, e.g., via recording the user's comments at home and replaying those during the visit. 

CAs could similarly provide rehabilitation support for social communication challenges, which is a common symptom among people with TBI and can seriously impair social relationships \cite{sohlberg2019part1, meulenbroek2019part2}. Participants reported being embarrassed talking in social groups because of their TBI. CAs could provide simulated practice of social communication skills (e.g., initiating conversation, turn-taking, using appropriate tone of voice) that can be reviewed with their speech-language pathologist. CAs can also function as an intermediary role in communications between people with TBI and their close others. For example, a CA can provide prompt questions or other suggestions (e.g., ``How about asking the person how their day went?'') and help structure communication interactions \cite{rietdijk2013training}. Furthermore, CAs can give reminders to do at-home rehabilitation practices, which would support long-term maintenance of rehabilitation gains after a user is discharged from therapy.

\subsubsection{CA Design for Routine Activities}
Participants have suggested areas of improving CAs to support their routine activities, such as activity logging, meal preparation, and daily routine. Individuals with TBI might need more time to start a task after being reminded by the CA, but they tend to forget afterwards due to short-term memory loss. In this case, a CA could check in with users again after a few minutes to ensure they have recalled key information or to prompt them to take appropriate action. The follow-up notifications can be essential and critical, e.g., for TBI users who can't smell and forget to turn off the stove. Besides reminders, future CAs could also be more effective in guiding users during multi-tasking and guiding them through completing tasks and switching tasks. When users' routine activities change due to unexpected events, CAs may keep track of the users' schedules and help users automatically re-plan activities.

\subsection{Limitations}
Our work also has a number of limitations that fall under two categories: (1) limitations due to system usability and limitations of the technology's capabilities; and (2) limitations due to study design and administration. We discuss these limitations below.

\subsubsection{Usability \& Technology.} First, some study participants faced usability challenges in setting up their systems, which diminished their experience with the conversational assistant. For example, setting up the location of the user was found to be difficult for some of the participants, which meant that the system could not provide location-based information or recommendations. Second, the conversational capabilities of smart speaker systems out of the box are highly limited, which limited the depth of social chat in which participants were engaged with the conversational assistant. Although we chose to study the baseline user experience with conversational assistants, future work can build custom applications that can provide more sophisticated and personalized conversational capabilities. Third, the audio recording tablets did not capture all the contextual information of the CA usage, which might have been caused by technical reasons and participants' privacy concerns (\S\ref{sec:data collection}). We acknowledged that the potential insights from missing audio could have been lost during the data collection process. 

\subsubsection{Study Design \& Administration.} Moreover, the study procedure limited the experience to being entirely naturalistic. Specifically, we set up the system with research accounts in order to facilitate data collection and to prevent privacy and security incidents that could arise from accessing participants' personal accounts for interaction logs. This study design choice limited the ability of the participant to use features and services that required a subscription or connect the system to other systems in their homes, such as smart home systems. 

Participants gave permission for us to collect their CA usage data and the contextual audio recordings before the weekly interviews. Knowing that the research team would be listening to the audio and checking the usage logs might have impacted the participants' CA use and the way they interacted with it. Although we found our study design to optimally balance our ability to collect extensive data and conduct a study in a naturalistic setting, future work can explore ways of conducting genuinely naturalistic data by recruiting users who have already adopted smart speaker systems and analyzing historical data. However, such approaches also may not be able to capture the usability and adoption issues surrounding these systems, as it will bias the study population to those who have not experienced them or those who have overcome such challenges. 

Furthermore, we focused on participants' CA usage during the four-week study period and did not have access to the usage data after the study. Nevertheless, the post-study usage data for a longer period can provide valuable insights for long-term CA use. Future work can conduct follow-up studies with the same group of participants to analyze their historical CA usage data to study the long-term usage pattern. 

Finally, our findings are based on data from a small number of individuals (nine). Recruiting and retaining individuals with TBI has a number of challenges, particularly for field and naturalistic studies, as the cognitive and social challenges experienced by this population may also limit their participation in research. Therefore, we chose to focus on depth of analysis over a large sample, although future work can follow a mixed-methods approach to combine technology probes, historical data, diary studies, and so on in order to capture a comprehensive picture of day-to-day use from a larger population.

\section{Conclusion}
Individuals with traumatic brain injury (TBI) suffer from cognitive, communication, and psychological challenges that profoundly limit their participation in everyday life. Conversational Agents (CAs) can help these individuals overcome challenges associated with these challenges through cognitive and communicative support, although much research is needed to understand how TBI users make use of CAs to address their injury-related needs, as well as the barriers to the adoption of CAs by the population. In this study, we provided nine adults in the chronic stage of moderate-severe TBI with a commercially available CA system and asked them to use it for four weeks to investigate use patterns, challenges, and design requirements, focusing particularly on injury-related uses. Our analysis of data from use logs, audio recordings during the use of the CA, and weekly interviews identified 14 activities that participants engaged in with CAs, that were related to TBI-related needs. These activities fell within four groups: cognitive health support, mental health support, cognitive activities, and routine activities. We also identified significant gaps between the current capabilities of CAs and the user needs and expectations in each activity, along with concrete design recommendations to address these issues. Our findings inform the design of future CA systems that can better support the cognitive, communicative, and psychological needs of users with TBI, as well as those who face similar cognitive challenges.
\section{Acknowledgement}
This work was funded by the National Institutes of Health (NIH R01-HD071089-06A1). Figure~\ref{fig:teaser} modified assets from Freepik.com for its design. We would like to thank our participants for their time and participation in this research study.

\balance
\bibliographystyle{ACM-Reference-Format}
\bibliography{literature}
\end{document}